\documentclass{aastex63}
\usepackage{natbib}

\begin{document}

\title{
Coherent Stellar Motion in Galactic Spiral Arms by Swing Amplification 
}

\shorttitle{Spiral Arms by Swing Amplification}
\shortauthors{Michikoshi \& Kokubo}

\author{Shugo Michikoshi}
\affiliation{ Department for the Study of Contemporary Society, Kyoto Women's University, Imakumano, Higashiyama, Kyoto, 605-8501, Japan }
\email{michikos@kyoto-wu.ac.jp}

\author{Eiichiro Kokubo}
\affiliation{Center for Computational Astrophysics, National Astronomical Observatory of Japan, Osawa, Mitaka, Tokyo 181-8588, Japan}
\email{kokubo.eiichiro@nao.ac.jp}

\begin{abstract}
  
We perform local $N$-body simulations of disk galaxies and investigate
 the evolution of spiral arms.
We calculate the time autocorrelation of the surface density of spiral
 arms and find that the typical evolution timescale is described by the
 epicycle period.  
We investigate the distribution of the orbital elements of stars and
 find that in spiral arms the epicycle motions of stars are in phase
 while the spatial distribution of the guiding center is nearly uniform.
These facts clearly show that the phase synchronization of the epicycle
 motion takes place, which is theoretically predicted by the swing
 amplification. 

\end{abstract}

\section{Introduction}

Galaxies with spiral arms are classified into three types: grand-design,
 multi-armed and flocculent galaxies.
One of the models to explain these spiral arms is swing amplification
 \citep{Goldreich1965, Julian1966, Toomre1981}. 
In a galactic disk, a density pattern rotates from leading to trailing
 due to shear.
If Toomre's $Q$ is $1\mbox{--}2$, a disk responses to small
 perturbations remarkably, in which the pattern amplitude can be
 significantly enhanced due to the self-gravity during rotation.
This mechanism is called swing amplification \citep{Toomre1981}. 
\cite{Goldreich1965} investigated the swing amplification with the
 hydrodynamic model. 
\cite{Julian1966} adopted the collisionless Boltzmann equation model and
 investigated the similar phenomenon. 
They found that with a perturber such as the corotating over-dense
 region trailing patterns are excited even though the disk is stable to
 the axisymmetric perturbations.  

The spirals generated by the swing amplification are not stationary but
 transient and recurrent, which appear and disappear continuously. 
This transient and recurrent picture is supported by $N$-body
 simulations for multi-arm spirals \citep{Sellwood1984, Toomre1991,
 Baba2009, Sellwood2000, Sellwood2010, Fujii2011, DOnghia2013}. 
Since the swing amplification model is constructed based on the linear and local approximations \citep{Julian1966, Toomre1981}, it is expected to be applicable to
 the spiral arms in multi-armed and flocculent galaxies.  

In the series of our works, we have investigated the role of the swing
 amplification in spiral arm formation by the analytical model and
 $N$-body simulations \citep{Michikoshi2014, Michikoshi2016, 
 Michikoshi2016a, Michikoshi2018} (Papers I, II, III, and IV). 
The recent researches suggest that the some aspects of the short-term
 activities cannot be explained only by the linear theory.
For example, $N$-body simulations showed that the overdense or underdense regions are formed by nonlinear interaction between transient spiral arms \citep{DOnghia2013, DOnghia2015, Kumamoto2016}. 
Nevertheless, the linear theory can capture some important aspects of the process. 
We have already confirmed that the quantitative predictions from the linear analytical model of the swing amplification agree well with $N$-body simulations (Paper I, II, and, IV).
A simple theoretical model of the swing amplification was proposed by
 \cite{Toomre1981}. 
Using this model, \cite{Michikoshi2016a} (Paper III) investigated the
 swing amplification process in detail. 
They pointed out that the phase synchronization of the stellar epicycle
 motion is a key process to understand the swing amplification. 
Regardless of the initial phases of the epicycle motion, the phases are
 synchronized as the spirals are amplified. 
However, the phase synchronization has not yet been confirmed in
$N$-body simulations. 
The goal of this paper is to clarify the phase synchronization in
 $N$-body simulations predicted in Paper III, which provides the 
 evidence of the swing amplification.  

\cite{Baba2013} performed the global $N$-body simulations and investigated the generation and destruction processes of spiral arms.
They extracted a typical spiral arm and analyzed the motion of stars in it.
They found that the swing amplification plays an important role in the formation and destruction of spiral arms.
We investigate the generation and destruction processes in local simulations in more detail.
The local $N$-body simulations can be directly compared with the model of swing amplification that is based on the local epicycle approximation \citep{Julian1966, Toomre1981}.
Furthermore in local $N$-body simulations, we can easily analyze the evolution of the orbital elements of particles.
This is helpful to understand the particle motion during the amplification process.

The outline of this paper is as follows.
In Section 2, we summarize the calculation method.
In Section 3, we present the results of simulations and show the phase
 synchronization due to the swing amplification. We examine the detailed amplification process.
Section 4 is devoted for a summary.

\section{Numerical Simulation}

We perform local $N$-body simulations of pure stellar disks with the
 epicycle approximation as in the previous works (Papers I and II). 
We briefly summarize the simulation method.
In contrast to global $N$-body simulations, in the local $N$-body
 simulation, we consider a small rotating region by employing a local
 shearing box \citep[e.g.,][]{Toomre1991, Fuchs2005}. 
Since we simulate only a part of the disk, we can perform high resolution
 simulations relatively easily.

We adopt a local Cartesian coordinate system ($x,y,z$), whose origin
 revolves around the galactic center with the circular frequency
 $\Omega$. 
The $x$-axis is directed radially outward, the $y$-axis is parallel to
 the direction of rotation, and the $z$-axis is normal to the $x$-$y$
 plane. 
We consider a small computational domain on the galactic midplane with
 the size $L_x$ and $L_y$, where $L_x$ and $L_y$ are the lengths in the $x$ and $y$ directions, respectively. 
The center of the computational box is located at the origin of the
 local Cartesian coordinate system. 
We assume that the box size is sufficiently shorter than the
 galactocentric distance, that is $L_x, L_y \ll a$ where $a$ is the
 galactocentric distance of the computational domain. 
In the epicycle approximation, we neglect the higher order terms
 with respect to $x$, $y$, and $z$, and obtain the approximated equation
 of motion for star $i$ as
\begin{eqnarray}
 \frac{\mathrm d^2 x_i}{\mathrm{d}t^2} &=& 2 \Omega \frac{\mathrm{d}
 y_i}{\mathrm{d}t} + \left(4 \Omega^2 - \kappa^2 \right) x_i + \sum_{j
 \ne i}^N \frac{G m (x_j - x_i)}{(r_{ij}^2+\epsilon^2)^{3/2}}, \\ 
 \frac{\mathrm d^2 y_i}{\mathrm{d}t^2} &=& - 2 \Omega \frac{\mathrm d
 x_i}{\mathrm{d}t} + \sum_{j \ne i}^N \frac{G m (y_j -
 y_i)}{(r_{ij}^2+\epsilon^2)^{3/2}},  \label{eq:eom} \\ 
 \frac{\mathrm d^2 z_i}{\mathrm{d}t^2} &=& - \nu^2 z_i + \sum_{j \ne i}^N
 \frac{G m (z_j - z_i)}{(r_{ij}^2+\epsilon^2)^{3/2}},  
\end{eqnarray}
 where $G$ is the gravitational constant, $m$ is the stellar mass,
 $r_{ij}$ is the distance between stars $i$ and $j$, and 
 $\kappa$ is the epicycle frequency
 \citep[e.g.,][]{Toomre1981, Toomre1991, Kokubo1992, Fuchs2005,
 Michikoshi2014, Michikoshi2016}. 
We assume that all stars have the same mass. 
The frequency $\nu$ is the frequency of the vertical motion and we adopt
 $\nu=3\Omega$.
The length scale $\epsilon$ is the softening parameter of the gravity
 and we adopt $\epsilon = r_\mathrm{t}/4$ 
 where $r_\mathrm{t}$ is the
 tidal radius of a star \citep[e.g.,][]{Kokubo1992, Michikoshi2014}
\begin{equation}
  r_\mathrm{t} = \left(\frac{2m G}{4 \Omega^2 - \kappa^2} \right)^{1/3}.
\end{equation}

We solve the equation of motion considering the shearing periodic
 boundary. 
The size of the computational domain $L_x$ and $L_y$ should be larger
 than the typical length scale of spiral arms. 
In this system, the typical length scale is the critical wavelength of
 the gravitational instability, \citep{Toomre1964}  
\begin{equation}
 \lambda_\mathrm{cr} = \frac{4 \pi^2 G \Sigma_0}{\kappa^2},
  \label{eq:lambda_cr}
\end{equation}
 where $\Sigma_0$ is the initial averaged surface density of stars. 
We adopt $L_x = L_y = L = 15 \lambda_\mathrm{cr}$.
The number of stars is $N=9.0 \times 10^5$. 

With the number of stars in $\lambda_\mathrm{cr}^2$, $N_\mathrm{c}$,
$\Sigma_0 = m N_\mathrm{c}/\lambda_\mathrm{cr}^2$.
In this paper, $N_\mathrm{c} = 4000$.
Substituting $\Sigma_0$ into Equation (\ref{eq:lambda_cr}) we obtain $\lambda_\mathrm{cr} =
(4 \pi^2 G m N_\mathrm{c}/\kappa^2)^{1/3}$. 
Thus, the ratio of the tidal radius to the critical wavelength is
\begin{equation}
  \frac{r_\mathrm{t}}{\lambda_\mathrm{cr}} = \left(\frac{1}{2\pi^2
  N_\mathrm{c}} \frac{\kappa^2}{4\Omega^2 - \kappa^2} \right)^{1/3}.
\end{equation}
This ratio depends on $\kappa/\Omega$ and takes $0.016 \mbox{--} 0.049$ for $\kappa/\Omega = 1.0 \mbox{--} 1.9$. 
The tidal radius is much shorter than the critical wavelength.

The initial radial velocity dispersion $\sigma_x$ is set so that
 the initial Toomre's $Q$ is $Q_\mathrm{ini} = 1.2$ where 
\begin{equation}
 Q_\mathrm{ini}= \frac{\sigma_x \kappa}{3.36 G \Sigma_0},
\end{equation}
 \citep{Toomre1964}.  
The epicycle frequency is a parameter.
We adopt $\kappa/\Omega =  1.0$ (model k0), $1.1$
 (model k1), $1.2$ (model k2), $\cdots$, $1.8$ (model k8). 
The shear rate is given by 
\begin{equation}
  \Gamma = 2 - \frac{\kappa^2}{2\Omega^2}.
\end{equation}
The fiducial model is model k4, whose shear rate is about $1.02$. 

Initially $x$ and $y$ positions of stars are distributed randomly.
The vertical distribution of stars is determined so that it is
consistent with the initial $Q$ value.

The equation of motion for each star is integrated using a second-order leapfrog integrator with time-step $\Omega \Delta t / 2\pi = 1/200$.
The self-gravity of stars is calculated using the special-purpose computer, GRAPE-DR \citep{Makino2007}. 

\section{Results}

\subsection{Lifetime of Spiral Arms}
First, we examine the typical evolution of structures.
Initially the surface density is almost uniform but includes the
small density fluctuation due to the particle noise. Thus, the density
fluctuation can grow by the self-gravity.
In any model, the density structures appear readily.  

\begin{figure}
   \plotone{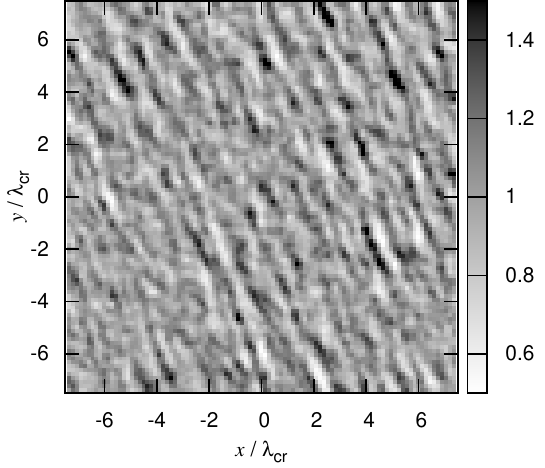}
 \caption{Snapshot of the surface density at $\Omega t / 2 \pi = 1.5$ for model k4. The surface density is normalized by $\Sigma_0$.}
  \label{fig:initial_snapshot}
\end{figure}

In the fiducial model (model k4), at $\Omega t / 2 \pi =0.5$, the trailing structures are generated spontaneously, which correspond to the spiral arms.
Figure \ref{fig:initial_snapshot} presents the surface density at $\Omega t / 2 \pi =1.5$.
We find the clear trailing structures with pitch angle of about $20^\circ$.
The separations between spiral arms in the $x$ and $y$ directions is roughly $\sim \lambda_\mathrm{cr}$ and $\sim 2\lambda_\mathrm{cr}$, respectively.
These results are consistent with the swing amplification model as shown in Papers I and II.

These spiral structures are not steady but transient and recurrent, which are generated and destroyed continuously.
This activity continues throughout the simulation time.
This behaviour has been observed in the local simulations \citep[][Paper I]{Toomre1991} and the global simulations \citep{Sellwood1984, Baba2009, Sellwood2000, Sellwood2010, Fujii2011}.  
Since the properties of these structures do not change during $\Omega t/ 2 \pi = 1.0 \mbox{--} 5.0 $, in the following we analyze the spiral arms during this period.

In order to analyze the averaged time-evolution of various quantities, at first we introduce the time average over $T$ of the space-time cross-correlation of quantity $f$ with $\Sigma$ as 
\begin{equation}
 \langle f \rangle (x,y,s) = \frac{1}{T \Sigma_0 L^2} \int \!\!\! \int
  \!\!\! \int f(x'+x,y'+y-2Ax's,t'+s) \Sigma(x',y',t') \mathrm{d}x' \mathrm{d}y' \mathrm{d}t', 
  \label{eq:crosscor}
\end{equation}
where $s$ is the lag for space-time cross-correlation. 
This function means the correlation between $\Sigma$ and $f$ at two different times and points, which traces the typical time evolution of $f$ around the overdense region.
It should be noted that in this formulation the shear motion is taken into account.
Because the averaged velocity of the focused region in the $y$ direction
 is given by the shear velocity $-2 A x'$, the typical displacement of the
 region in the $y$ direction during time $s$ is expected to be $-2 A x' s$, where $A$ is Oort's $A$ constant. Thus we introduce the offset $-2 A x' s$ into the $y$ component.

Choosing $f=\Sigma/\Sigma_0 - 1$, we define the space-time autocorrelation as
\begin{equation}
  \eta (x,y,s) = \left\langle \frac{\Sigma}{\Sigma_0} - 1 \right\rangle (x,y,s),
 \label{eq:spatialautocor}
\end{equation}
which shows the typical evolution of the surface density fluctuation around the overdense region.
Setting $x=y=0$, we obtain the time autocorrelation function as
\begin{equation}
 \bar \Psi (s) =  \eta (0,0,s),
 \label{eq:timeautocor}
\end{equation}
 which means the typical time evolution at the center of the overdense region.

\begin{figure}
 \begin{center}
  \plotone {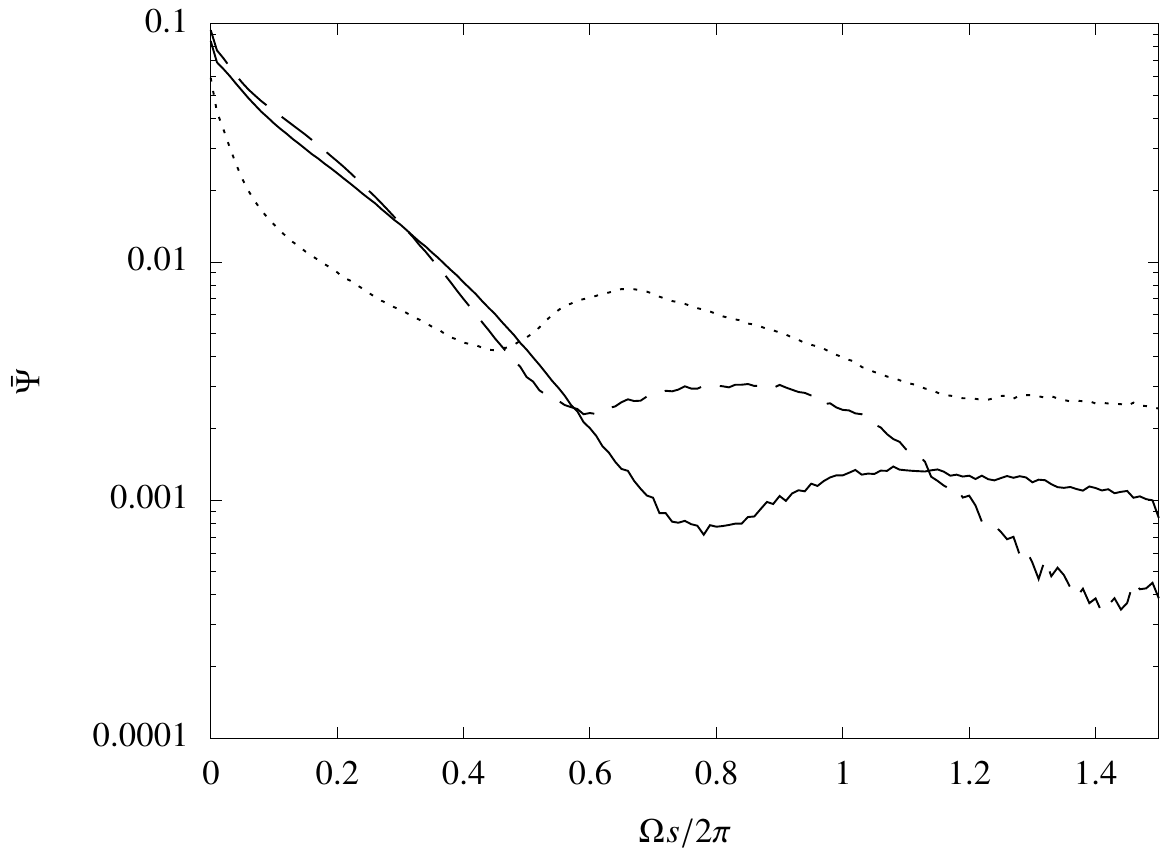}
 \end{center}
\caption{
Averaged time autocorrelation function of the surface density $\bar\Psi$ for $\kappa / \Omega=1.0$ (model k0) (solid), $1.4$ (model k4) (dashed) and $1.8$ (model k8) (dotted).
} 
 \label{fig:timecore}
\end{figure}

The time autocorrelation functions for models k0, k4, and k8 are shown
 in Figure \ref{fig:timecore}. 
The time autocorrelation decreases with increasing $s$.
This means that the overdense region declines with time.
We find the local minimum and the local maximum.
On average the density at an overdense region tends
 to increase again after its first decay, which seems to be a damped oscillation \citep{Julian1966}.

\begin{figure}
 \begin{center}
  \plotone {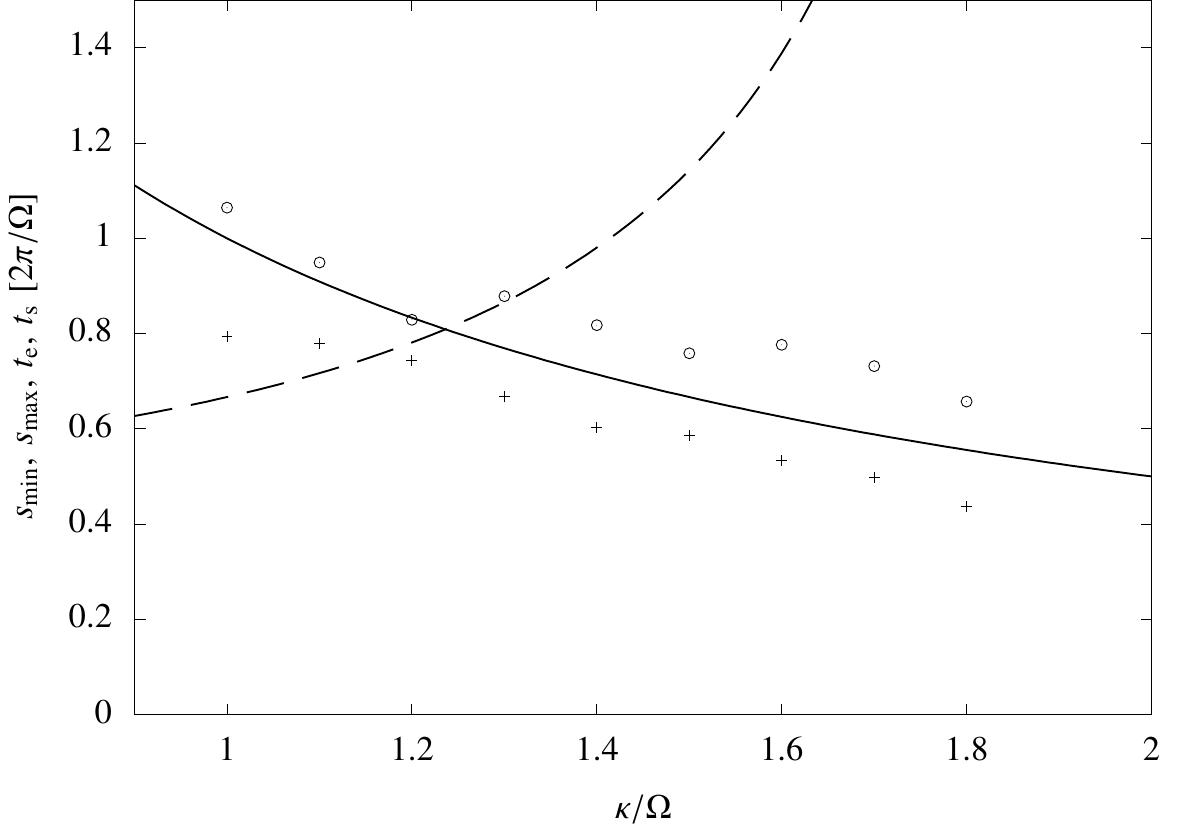}
 \end{center}
\caption{
Local minimum time $s_\mathrm{min}$ (plus) and local maximum time $s_\mathrm{max}$ (circle) of the time autocorrelation function as a
 function of $\kappa / \Omega$.
   The solid and dashed curves show the epicycle period $t_\mathrm{e}$ and
  the shear time $t_\mathrm{s}$, respectively.
} 
 \label{fig:decaytime}
\end{figure}

The damping time of the time autocorrelation function is the typical
 timescale of the activity of spiral arms. 
We define $s_\mathrm{min}$ as the time when $\bar \Psi$ takes the first local minimum. 
Similarly we define $s_\mathrm{max}$ as the time when $\bar \Psi$ reaches the local maximum after $s_\mathrm{min}$. 
The dependencies of $s_\mathrm{min}$ and $s_\mathrm{max}$ on
 $\kappa/\Omega$ are summarized in Figure \ref{fig:decaytime}. 
Both $s_\mathrm{min}$ and $s_\mathrm{max}$ decrease with increasing
 $\kappa/\Omega$. 
We compare them with two dynamical timescales, the epicycle period
 $t_\mathrm{e} = 2\pi / \kappa$ and the shear timescale $t_\mathrm{s} = 1/(2A)$.
If the spiral arms are destroyed by the shear or the tidal force, it
 is expected that the damping time is characterized by the shear timescale. 
However, the shear timescale increases with $\kappa / \Omega$, and its
 dependence on $\kappa / \Omega$ is completely different from those of
 $s_\mathrm{min}$ and $s_\mathrm{max}$. 
On the other hand, the epicycle period decreases with increasing
 $\kappa/\Omega$, which has a similar dependence to $s_\mathrm{min}$ and
 $s_\mathrm{max}$. 
Thus the epicycle motion relates to generation and destruction processes.

The time autocorrelation function evolves like a damped oscillation.  
This is consistent with the swing amplification model discussed in Paper III \citep{Julian1966}.
The elementary process of the swing amplification is the phase synchronization of the epicycle motion.
Therefore, the timescale of the spiral activity is also described by
 the epicycle period.  

\subsection{Phase Synchronization of Epicycle Motion \label{sec:phasesync}}

If we neglect the self-gravity of stars, the motion of a star
 is separated into two components: a guiding center and an epicycle,
 which are given as \citep[e.g.,][]{Binney2008}
\begin{eqnarray}
 x &=& x_\mathrm{g} - x_\mathrm{a} \cos \phi, \label{eq:eomx} \\
 y &=& y_\mathrm{g} + \frac{2 x_\mathrm{a} \Omega}{\kappa} \sin \phi, \label{eq:eomy} 
  \label{eq:epicycle}
\end{eqnarray}
 where $(x_\mathrm{g}, y_\mathrm{g})$ is the position of the guiding
 center, $x_\mathrm{a}$ is the amplitude of the epicycle oscillation, and $\phi$ is its phase.
The $x$ component of the guiding center $x_\mathrm{g}$ remains constant while its $y$ component $y_\mathrm{g}$ is given as
\begin{equation}
 y_\mathrm{g}= -2A t x_\mathrm{g} + y_\mathrm{g0},
 \label{eq:guidey}
\end{equation}
 where $y_\mathrm{g0}$ is the initial $y$ component of the guiding center.  
 The phase $\phi$ varies with time as 
\begin{equation}
  \phi = \kappa t - \phi_0,
\end{equation}
where $\phi_0$ is the initial phase.
Using Equations (\ref{eq:eomx}), (\ref{eq:eomy}) and (\ref{eq:guidey})
 we can calculate $x_\mathrm{g}, y_\mathrm{g}, \phi$ and
 $x_\mathrm{a}$ from the position $(x,y)$ and 
 velocity $(\mathrm{d}x/\mathrm{d}t, \mathrm{d}y/\mathrm{d}t)$. 

We divide the computational domain into $150 \times 150$ cells.
Selecting the stars whose guiding centers are in each cell, we
 calculate the average of the relative position of stars to their
 guiding centers $(x-x_\mathrm{g}, y-y_\mathrm{g})$ and the
 corresponding phase $\bar \phi $ from
\begin{equation}
 \tan \bar \phi = -  \frac{\kappa}{2 \Omega} \frac{\langle y-y_\mathrm{g} \rangle}{\langle x-x_\mathrm{g} \rangle },
\end{equation}
 where angle brackets denote the average in each cell.

\begin{figure}
 \begin{minipage}{0.07\hsize}
   \begin{flushright}
   $\Sigma$
   \end{flushright}
 \end{minipage}
 \begin{minipage}{0.3\hsize}
 \begin{center}  $\Omega t / 2 \pi = 0.0$ \end{center}
   \includegraphics[width=\columnwidth]{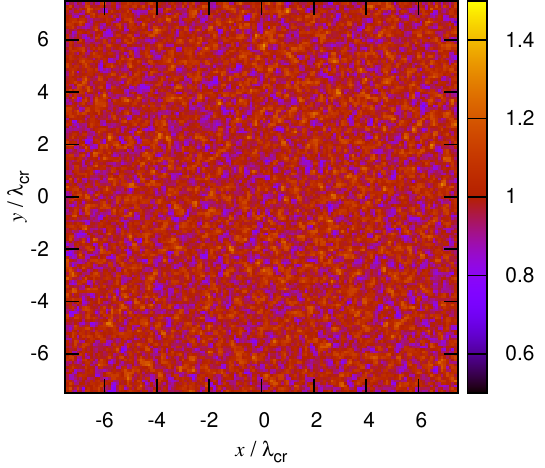}
 \end{minipage}
 \begin{minipage}{0.3\hsize}
   \begin{center} $\Omega t / 2 \pi = 2.0$ \end{center}
   \includegraphics[width=\columnwidth]{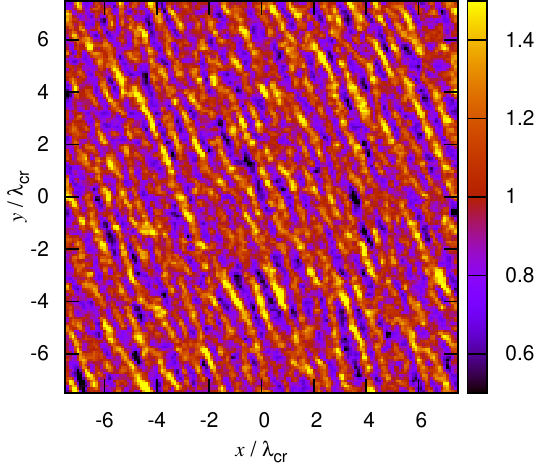}
 \end{minipage}
 \begin{minipage}{0.3\hsize}
 \begin{center}  $\Omega t / 2 \pi= 4.0$ \end{center}
   \includegraphics[width=\columnwidth]{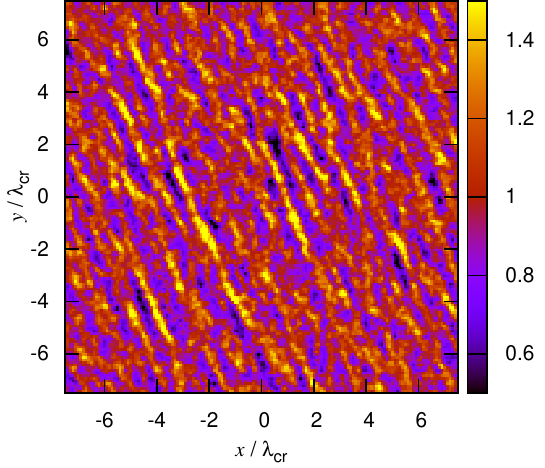}
 \end{minipage}

 \begin{minipage}{0.07\hsize}
   \begin{flushright}
     $\Sigma_\mathrm{g}$
   \end{flushright}
 \end{minipage}
 \begin{minipage}{0.3\hsize}
   \includegraphics[width=\columnwidth]{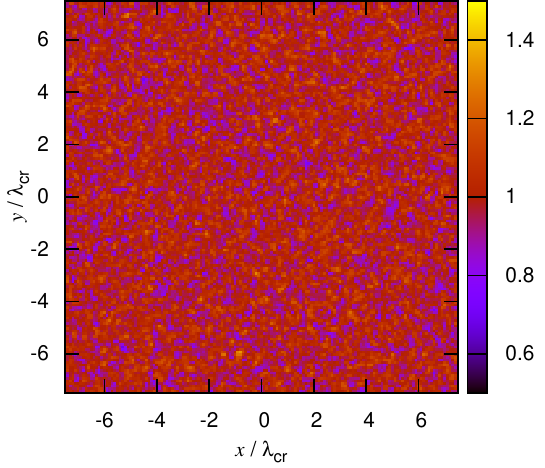}
 \end{minipage}
 \begin{minipage}{0.3\hsize}
   \includegraphics[width=\columnwidth]{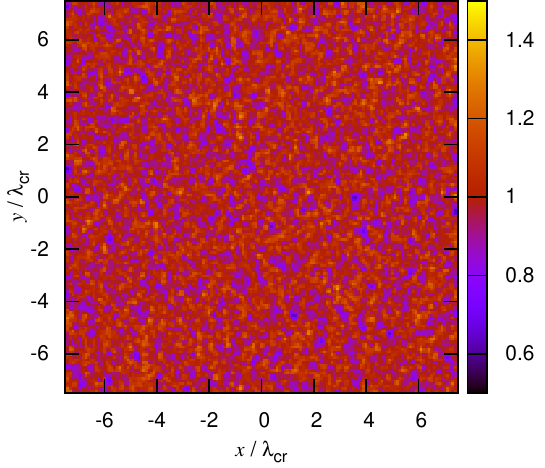}
 \end{minipage}
 \begin{minipage}{0.3\hsize}
   \includegraphics[width=\columnwidth]{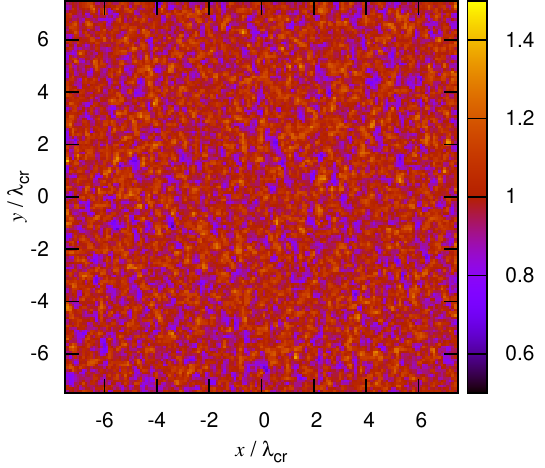}
 \end{minipage}

 \begin{minipage}{0.07\hsize}
   \begin{flushright}
   $\bar \phi$
   \end{flushright}
 \end{minipage}
 \begin{minipage}{0.3\hsize}
   \includegraphics[width=\columnwidth]{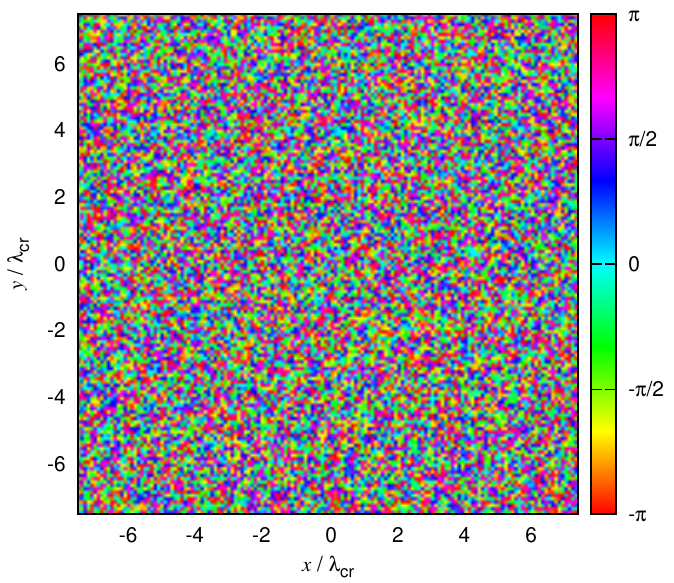}
 \end{minipage}
 \begin{minipage}{0.3\hsize}
   \includegraphics[width=\columnwidth]{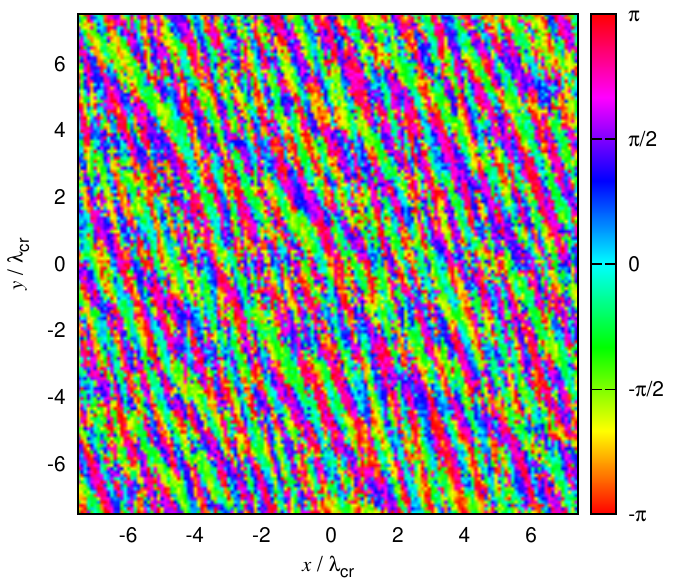}
 \end{minipage}
 \begin{minipage}{0.3\hsize}
   \includegraphics[width=\columnwidth]{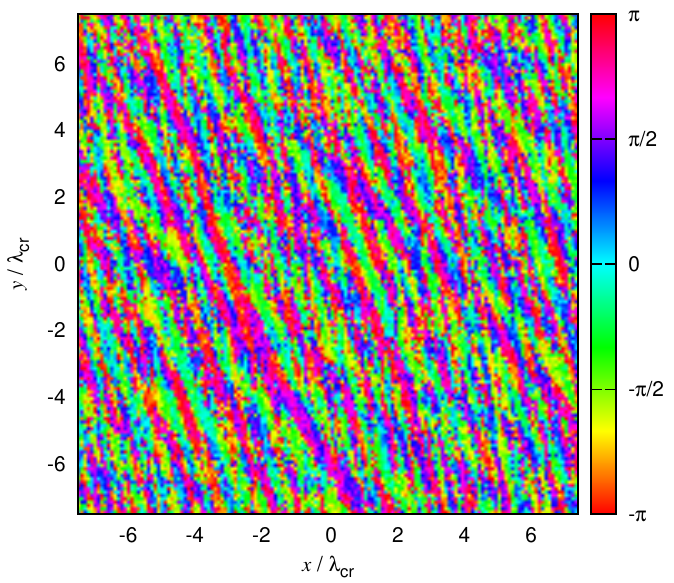}
 \end{minipage}

\caption{
Surface densities of stars $\Sigma$ (top) and their guiding center
 $\Sigma_\mathrm{g}$ (middle) normalized by $\Sigma_0$ and the epicycle
 phase $\bar \phi$ (bottom) at $\Omega t / 2 \pi= 0.0$ (left), $\Omega t / 2 \pi= 2.0$
 (middle), and $\Omega t / 2 \pi= 4.0$ (right) for model k4. 
}
\label{fig:snapshot_evo}
\end{figure}

Figure \ref{fig:snapshot_evo} shows the surface densities of stars
 $\Sigma$ and their guiding centers $\Sigma_\mathrm{g}$ and $\bar \phi$. 
At $t=0$, because the stars are distributed uniformly, in other
 words, their guiding centers and epicycle phases are given randomly,
 $\Sigma$, $\Sigma_\mathrm{g}$ and $\bar \phi$ have no structure
 completely.    
After that, the gravitational instability takes place and the spatial 
 structure appears. 
At $\Omega t / 2 \pi = 2.0$, $\Sigma$ and $\bar \phi$ show trailing structures, while
 $\Sigma_\mathrm{g}$ is nearly uniform.
At $\Omega t / 2 \pi = 4.0$, basically the structure remains the same. 
These results clearly show that the phase synchronization of the
epicycle motion enhances the surface density of stars in the
spiral arms. During the phase synchronization the spatial distribution
of the guiding centers is kept almost uniform since their change is not significant.
 
\clearpage

\subsection{Stellar Motion in Spiral Arms}
The stellar motion in spiral arms is important to understand generation and destruction processes of spiral arms \citep{Baba2013}.
We extract a typical spiral arm and investigate dynamics of stars in it.

\begin{figure}
 \begin{minipage}{0.49\hsize}
   \begin{center} (A) $\Omega t/2\pi = 3.6$ \end{center}
     \vspace{-14pt}
   \includegraphics[width=\columnwidth]{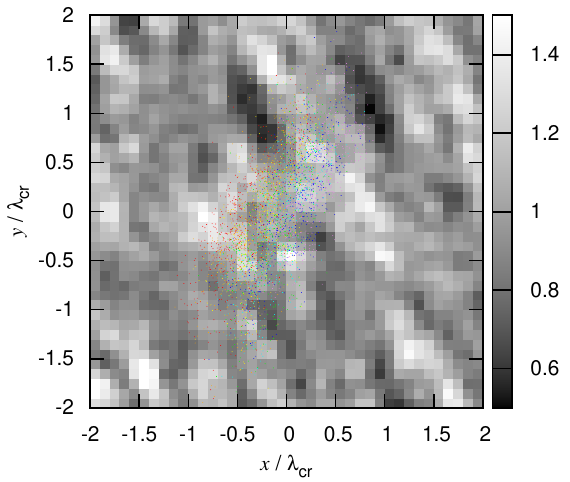}
 \end{minipage}
 \begin{minipage}{0.49\hsize}
 \begin{center} (B) $\Omega t/2\pi = 3.8$ \end{center}
     \vspace{-14pt}
   \includegraphics[width=\columnwidth]{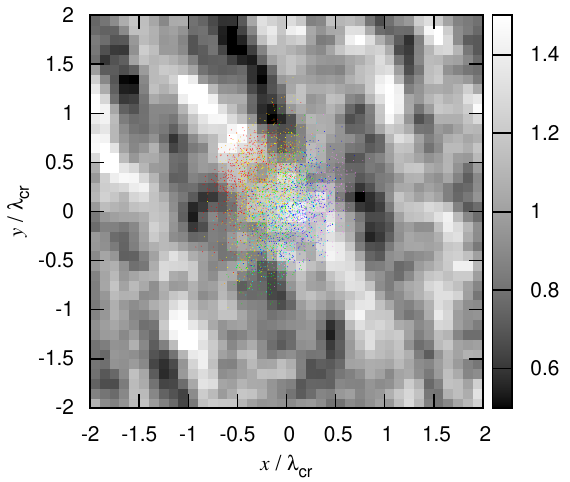}
 \end{minipage}
 \begin{minipage}{0.49\hsize}
   \vspace{15pt}  
 \begin{center}  (C) $\Omega t/2\pi = 4.0$ \end{center}
     \vspace{-14pt}
   \includegraphics[width=\columnwidth]{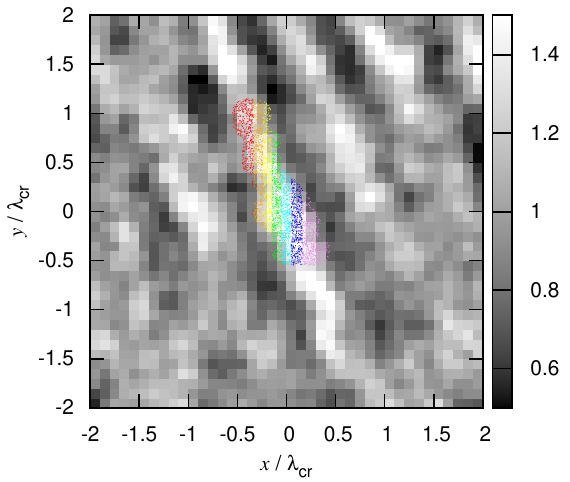}
 \end{minipage}
 \begin{minipage}{0.49\hsize}
   \vspace{15pt}  
   \begin{center}  (D) $\Omega t/2\pi = 4.4$ \end{center}
     \vspace{-14pt}
   \includegraphics[width=\columnwidth]{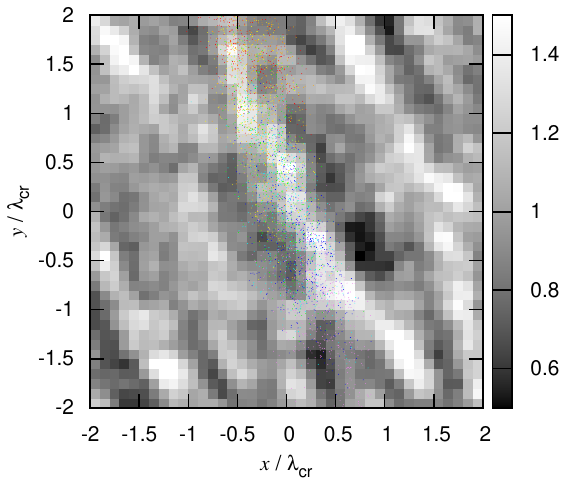}
 \end{minipage}
 \caption{Stars in the amplified spiral arms at $\Omega t/2\pi  = 3.6$, $3.8$, $4.0$, and $4.4$ for model k4.
   The color shows groups classified by their position at $\Omega t/2\pi  = 4.0$. The position $x$ and $y$ is relative position to the center of the focusing spiral.
The grey scale map denotes the surface density normalized by $\Sigma_0$. 
  \label{fig:typical}
}
\end{figure}

We search the highest surface density cell at $\Omega t / 2\pi=4.0$ for model k4, which is located at $(x,y)=(1.25 \lambda_\mathrm{cr}, -0.75 \lambda_\mathrm{cr})$. 
Next we extract the group of the high density cells with $\Sigma/\Sigma_0>1.4$ that include the highest surface density cell and connect to each other.
These cells consist of the amplified spiral arms. 
We investigate the motion of stars in this region.
Figure \ref{fig:typical}C shows these stars at $\Omega t / 2 \pi=4.0$.
The stars are separated into 7 groups by their $x$ position and we distinguish them by color.

Figure \ref{fig:typical} shows the spatial distribution of the selected stars at $\Omega t/2 \pi=3.6 \mbox{--} 4.4$.  
At $\Omega t/2 \pi = 3.6$ (Fig. \ref{fig:typical}A), the stars are scattered in the leading pattern.
Although they diffuse in some degree, we can see the coherency of stars.
At $\Omega t/2 \pi = 3.8$ (Fig. \ref{fig:typical}B), they come close to the center.
The width in the $x$ direction becomes small.
At $\Omega t/2 \pi = 4.0$ (Fig. \ref{fig:typical}C), they concentrates on the spiral arm, which show the clear trailing pattern.
At $\Omega t/2 \pi = 4.4$ (Fig. \ref{fig:typical}D), the width of the pattern widens and the density decreases finally. 
During the rotation of the pattern from leading to trailing, the density is enhanced.
The rotation of the pattern and the density enhancement coincide, which is consistent with the swing amplification mechanism.

\begin{figure}
  \plotone{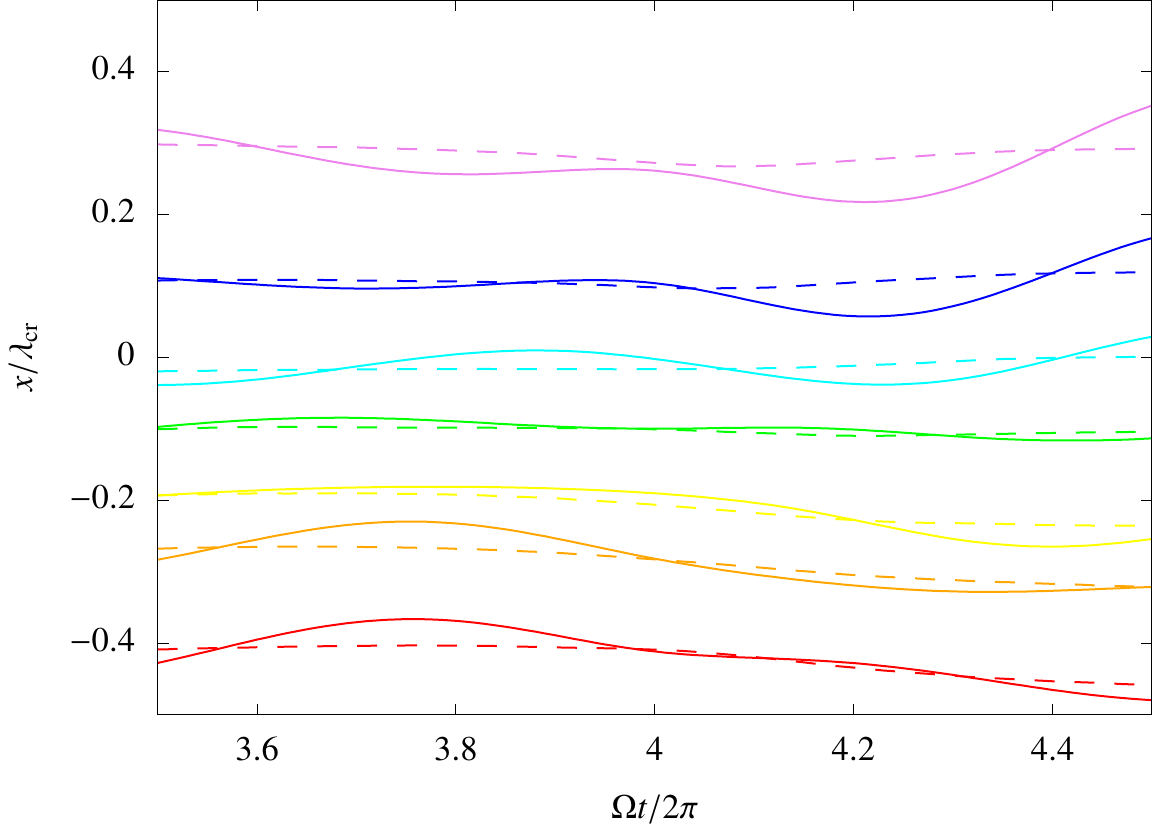}
  \caption{Mean $x$ position of the selected stars (solid) and the average $x$ position of their guiding centers (dashed).
  The color shows groups as in Figure \ref{fig:typical}.
  }
  \label{fig:txg}
\end{figure}

\begin{figure}
  \plotone{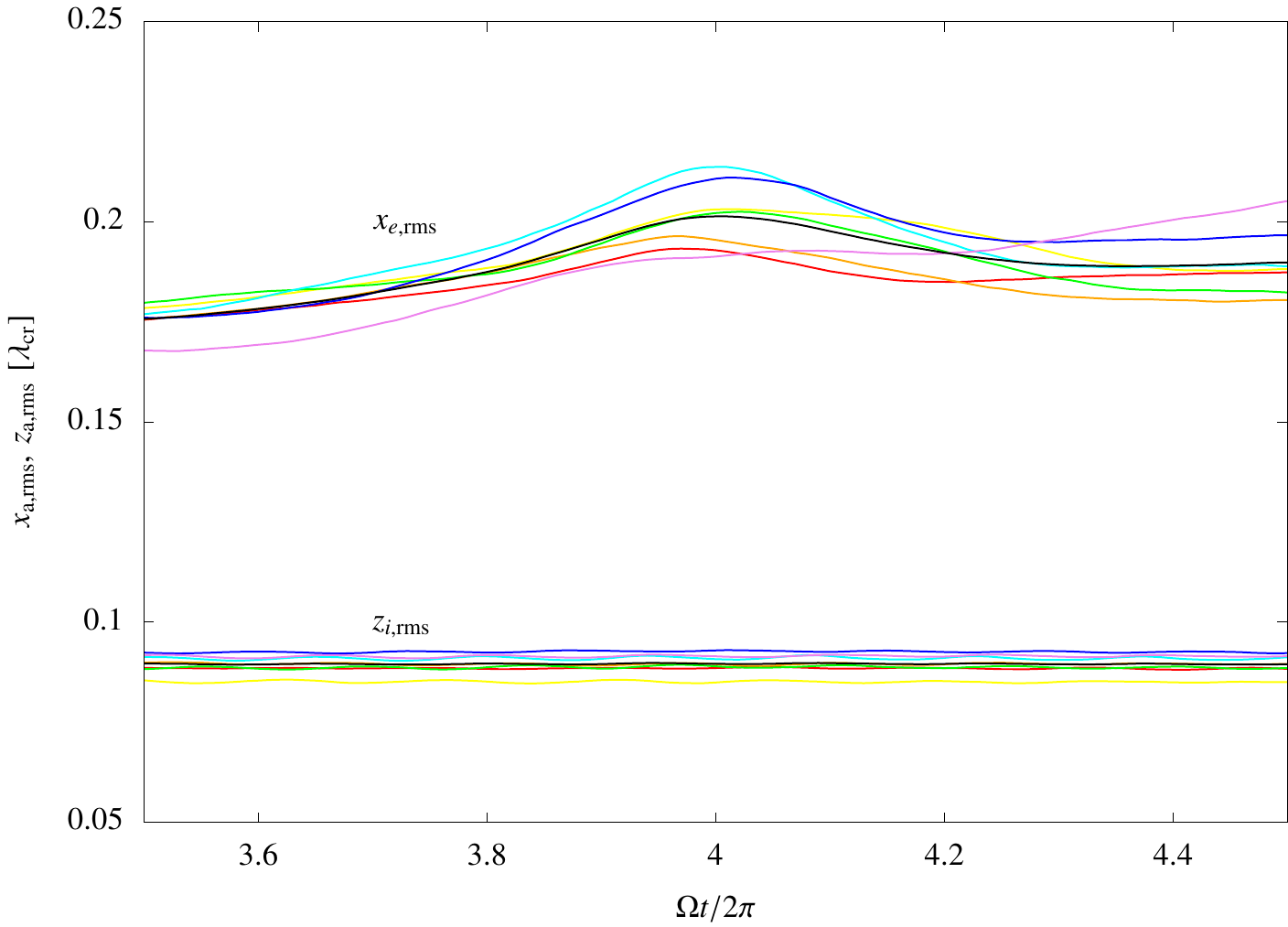}
  \caption{Time evolution of the root mean squares of the epicycle and vertical motions $x_{\mathrm{a},\mathrm{rms}}$ and $z_{\mathrm{a},\mathrm{rms}}$.
  The color shows groups as in Figure \ref{fig:typical}.
  The black curve shows the average of all groups.
  }
  \label{fig:txgamp}
\end{figure}

Figure \ref{fig:txg} shows the evolution of the averaged $x$ position of the stars and their guiding centers of each group.
The variations of the guiding centers are smaller than those of their positions.
This is consistent with the fact that the guiding center distribution remains uniform although the spiral arms are generated as discussed in Section \ref{sec:phasesync}.
To clarify the amplification, we examine the evolution of the amplitudes of the epicycle and vertical motion. 
The amplitude of the epicycle motion $x_\mathrm{a}$ is defined in Equation (\ref{eq:epicycle}), and that of the vertical motion $z_\mathrm{a}$ is defined by
\begin{equation}
  z = z_\mathrm{a} \cos (\nu t + \psi_0), \label{eq:eomz} 
\end{equation}
where $\psi_0$ is the phase of the vertical motion at $t=0$.
We introduce $x_{\mathrm{a},\mathrm{rms}}$ and $z_{\mathrm{a},\mathrm{rms}}$, that are the root mean squares of $x_\mathrm{a}$ and $z_\mathrm{a}$, respectively.
Figure \ref{fig:txgamp} displays the time evolution of $x_{\mathrm{a},\mathrm{rms}}$ and  $z_{\mathrm{a},\mathrm{rms}}$.
We find that is the largest $x_{\mathrm{a},\mathrm{rms}}$ around $\Omega t/2\pi = 4$.
Thus the density enhancement and the increase of the epicycle amplitude coincide.
This is consistent with the swing amplification model.
On the other hand, the $z_{\mathrm{a},\mathrm{rms}}$ barely changes during the amplification.
In the swing amplification model, the motion in the $z$ direction is not considered.
The numerical simulation shows that this treatment is valid. 
The swing amplification is essentially two-dimensional phenomenon.

\begin{figure}
  \plottwo{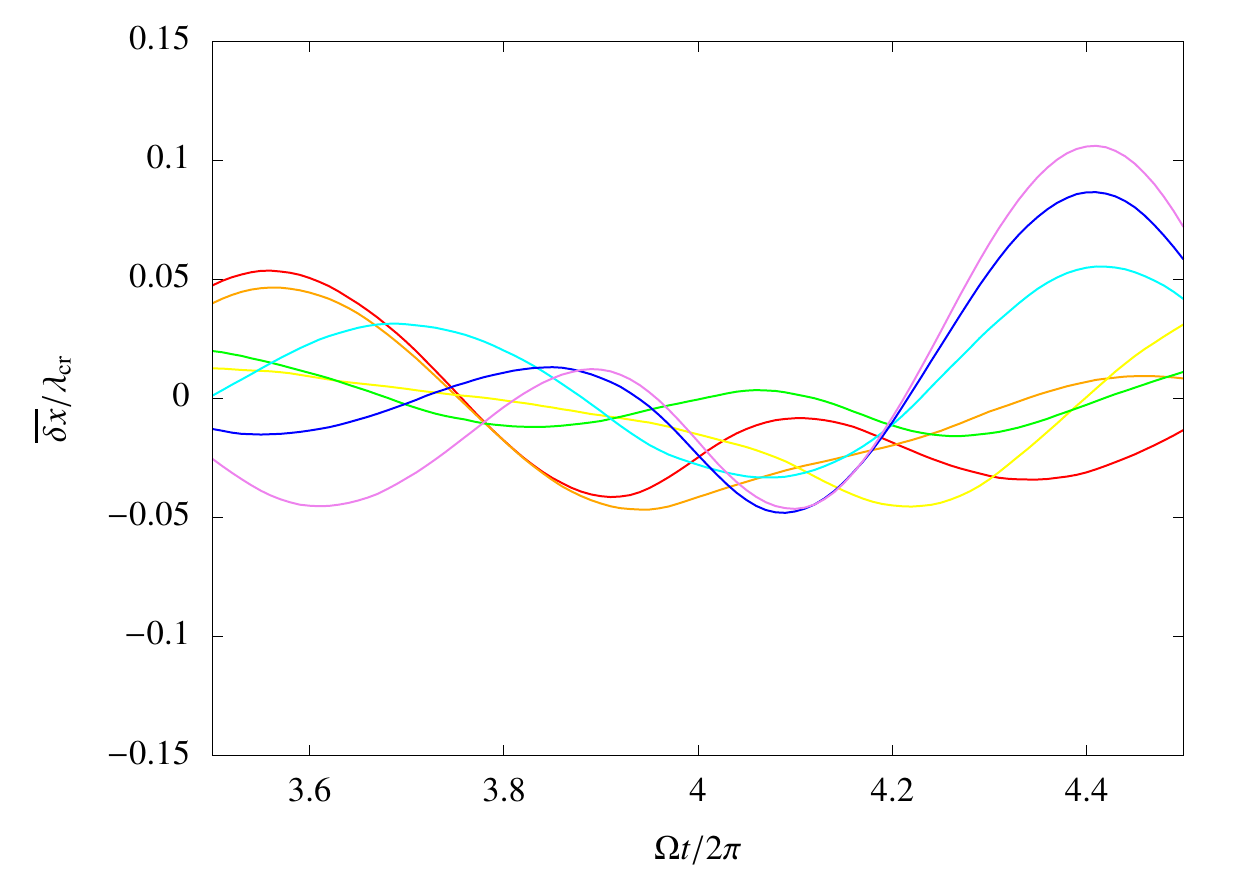}{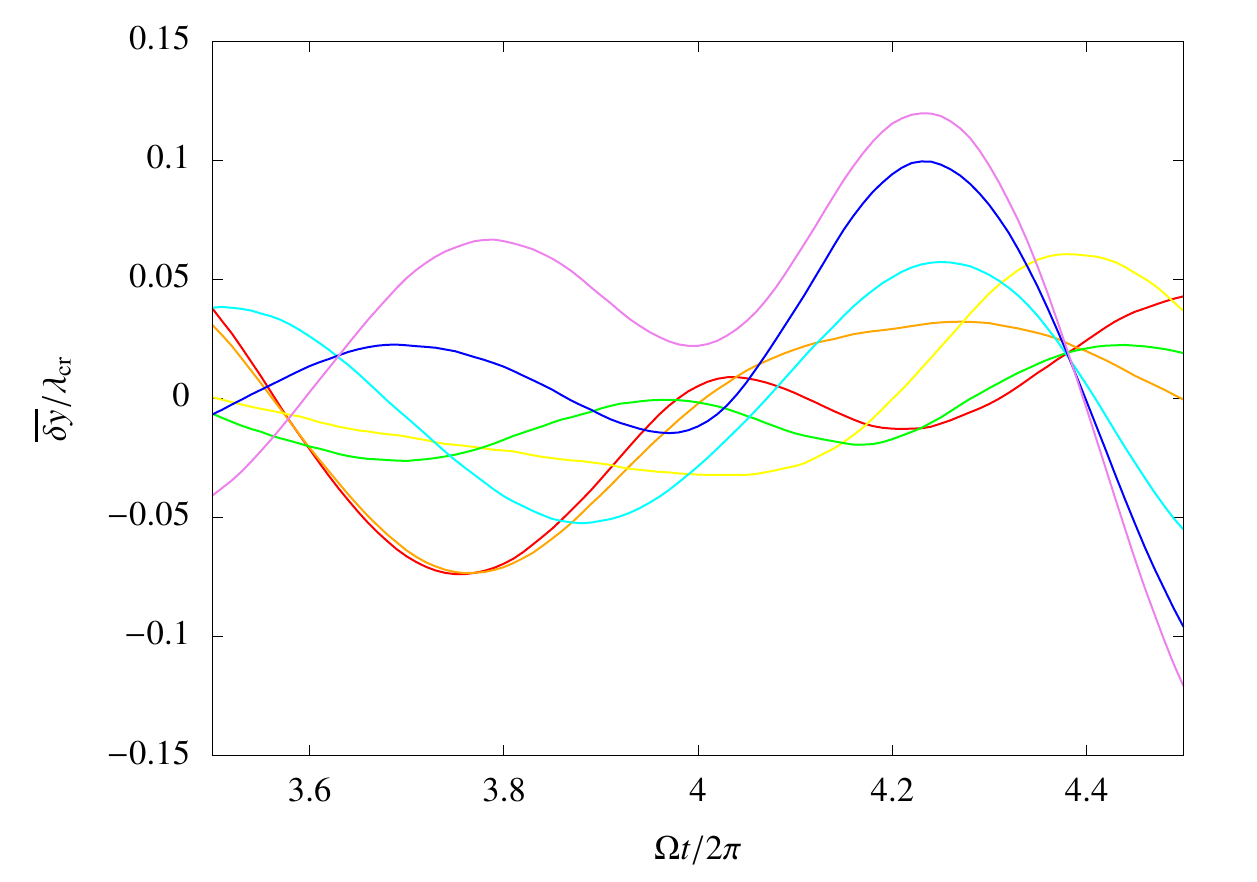}
  \caption{Time evolution of $\overline{\delta x}$ (left panel) and $\overline{\delta y}$ (right panel).
  The color shows groups as in Figure \ref{fig:typical}.
  }
  \label{fig:tdev}
\end{figure}

To examine the phase synchronization, we consider the displacement from the guiding center $\delta x = x - x_\mathrm{g}$ and $\delta y= y - y_\mathrm{g}$.
If the epicyclic oscillation is uniform, the average of $\delta x$ and $\delta y$ should be zero.
We define $\overline{\delta x}$ and $\overline{\delta y}$ as the average of $\delta x$ and $\delta
y$ in each group, respectively. The absolute values of these quantities show
the degree of the phase synchronization. If the epicycle phases of
stars are not synchronized, these are close to $0$. 
In Figure \ref{fig:tdev}, some groups show the upward trend while the other groups shows the downward trend.
Before the density is amplified ($\Omega t/2\pi < 3.8 $),
$|\overline{\delta x}| \lesssim 0.05$ and $|\overline{\delta y}| \lesssim 0.07$. 
Thus, the phase in each group is not well synchronized.
After the density amplification ($\Omega t/2\pi > 4 $), 
$|\overline{\delta x}|$ and $|\overline{\delta y}|$ in each
group increase up to 0.13. Thus, the phase in each group is synchronized
after the amplification.

As discussed above the stars gather in the $x$ direction and the width of the spiral arm shrinks in Figure 5 when the density gets enhanced ($\Omega t/2\pi = 3.8 \mbox{--} 4.2$).
This behavior cannot be understood by the previous analytical works based on the linear theory \citep[][Paper IV]{Toomre1981}.
In their analytical works, the density enhancement is caused by the displacement normal to a spiral arm.
Thus, the averaged displacement of all stars parallel to a pattern is zero. 
This discrepancy between the simulations and the analytical analyses suggests the importance of the finite length of spiral arms.

\subsection{Formation and Destruction of Spiral Arms}

\begin{figure}
 \begin{minipage}{0.33\hsize}
 \begin{center} $\Omega s/2 \pi = -0.90$ \end{center}
     \vspace{-14pt}
   \includegraphics[width=\columnwidth]{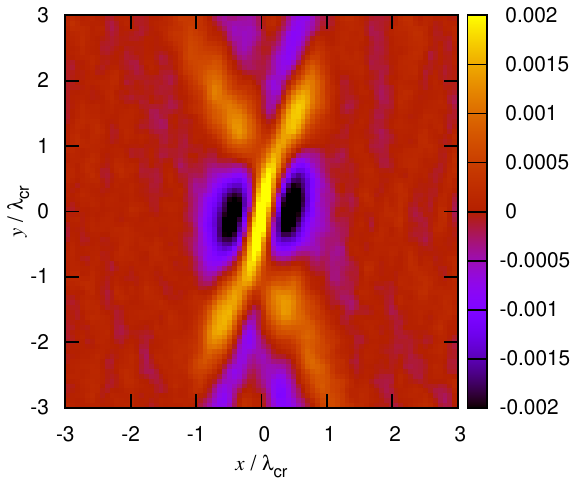}
 \end{minipage}
 \begin{minipage}{0.33\hsize}
 \begin{center} $\Omega s/2 \pi = -0.60$ \end{center}
     \vspace{-14pt}
   \includegraphics[width=\columnwidth]{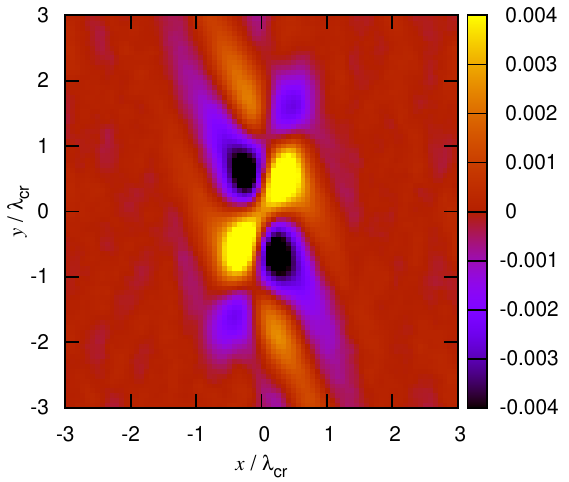}
 \end{minipage}
 \begin{minipage}{0.33\hsize}
 \begin{center} $\Omega s/2 \pi = -0.30$ \end{center}
     \vspace{-14pt}
   \includegraphics[width=\columnwidth]{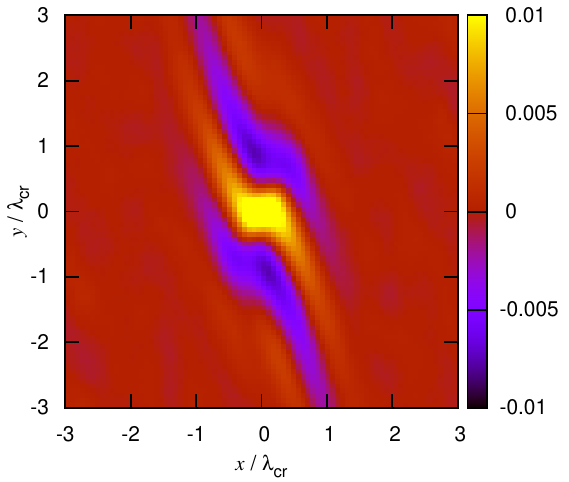}
 \end{minipage}

 \begin{minipage}{0.33\hsize}
   \vspace{13pt}
 \begin{center} $\Omega s/2 \pi = 0.00$ \end{center}
     \vspace{-14pt}
   \includegraphics[width=\columnwidth]{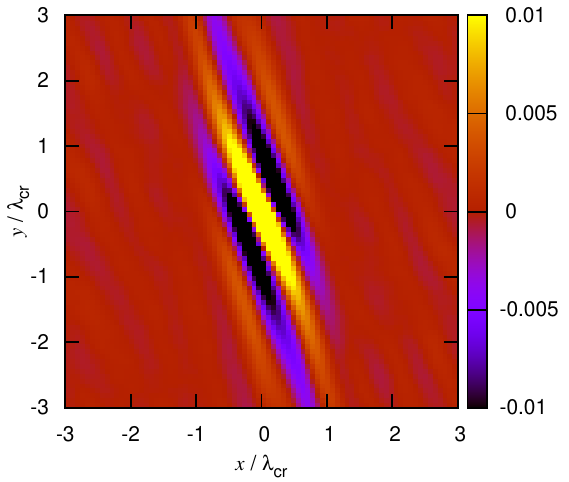}
 \end{minipage}
 \begin{minipage}{0.33\hsize}
   \vspace{13pt}
 \begin{center} $\Omega s/2 \pi = 0.30$ \end{center}
     \vspace{-14pt}
   \includegraphics[width=\columnwidth]{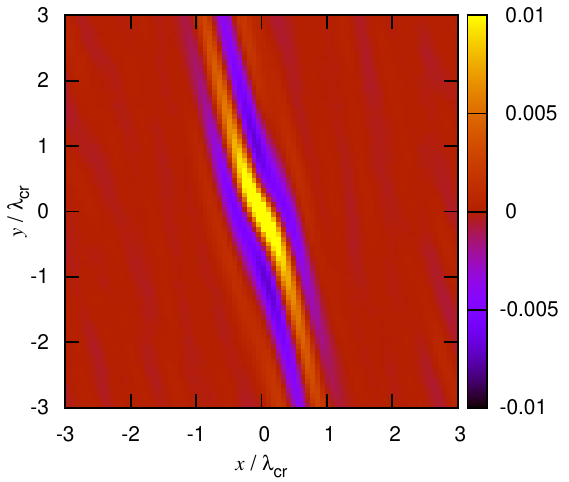}
 \end{minipage}
 \begin{minipage}{0.33\hsize}
   \vspace{13pt}
 \begin{center} $\Omega s/2 \pi = 0.60$ \end{center}
     \vspace{-14pt}
   \includegraphics[width=\columnwidth]{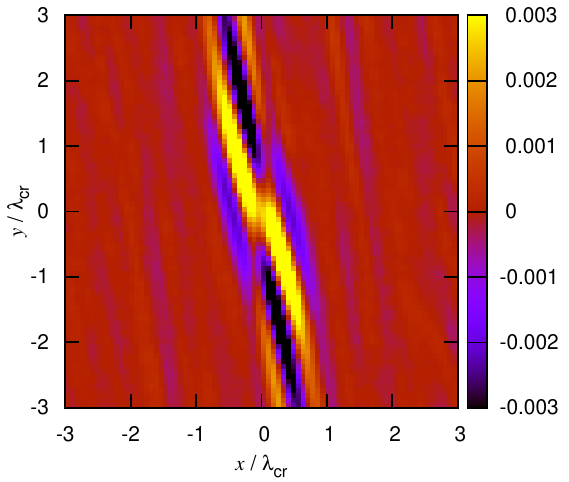}
 \end{minipage}
\caption{
  Space-time autocorrelation function $\eta$ at $\Omega s/2\pi = -0.90$, $-0.60$, $-0.30$, $0.00$, $0.30$, and $0.60$ for model k4 (from top left to bottom right).
}
\label{fig:delyed_evolution}
\end{figure}

In order to analyze the typical spiral evolution, we use the space-time autocorrelation given by Equation (\ref{eq:spatialautocor}). 
Figure \ref{fig:delyed_evolution} shows the space-time autocorrelation
 function at $\Omega s/2 \pi = -0.90, -0.60, -0.30, 0.00, 0.30,$ and $0.60$.
At $\Omega s/2 \pi = -0.90$, we can observe the faint leading structure. 
The overdense region has the leading structure before the density
 enhancement.
Because of the shear, the pitch angle increases gradually.
At $\Omega s/2 \pi = -0.60$, there are two overdense regions at the both ends
 of the leading structure. 
Each overdense region has the dim trailing tails.
At $\Omega s/2 \pi = -0.30$, although the pitch angle at the center is quite
 large, it has already trailing tails.
At $\Omega s/2 \pi = 0.0$, the spiral arms are amplified to the maximum, and
 the structure is almost along the line. 
After the amplification, the amplitude begins to decrease.
At $\Omega s/2 \pi = 0.3$, the spiral structure bends at the center, that is,
 the pitch angle at the center is larger than those in the tails. 
At $\Omega s/2 \pi = 0.6$, the spiral structure splits into two halves and
 the narrow leading structure develops.  

The amplification processes in the numerical simulation and the analytical analyses share the basic fact that the amplification occurs while the
 pattern rotates from leading to trailing. 
However, the overall process depicted here is somewhat different from
 the behavior considered in the analytical analyses. 
In the analytical analyses, we consider only the rotating single wave and do not consider the structure parallel to the wave. 
It seems that the leading pattern forms from the interaction between two
 trailing spiral arms. 

\begin{figure}
\begin{minipage}{0.49\hsize}
\begin{center} $\Omega s/2\pi = -0.40$ \end{center}
 \vspace{-14pt}
 \plotone{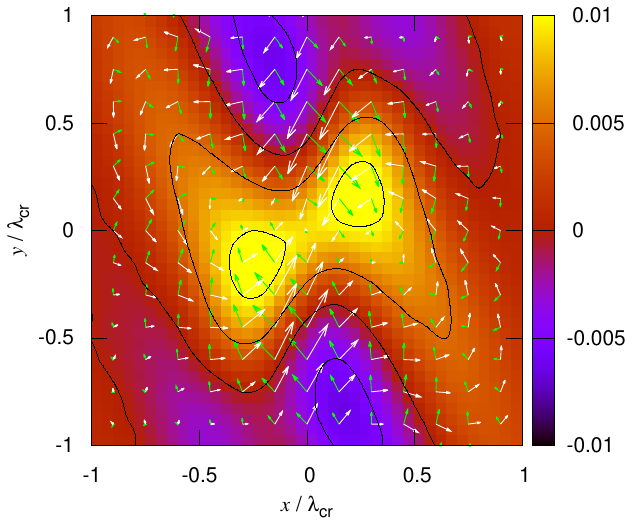} 
\end{minipage}
\begin{minipage}{0.49\hsize}
\begin{center} $\Omega s/2\pi = -0.20$ \end{center}
 \vspace{-14pt}
 \plotone{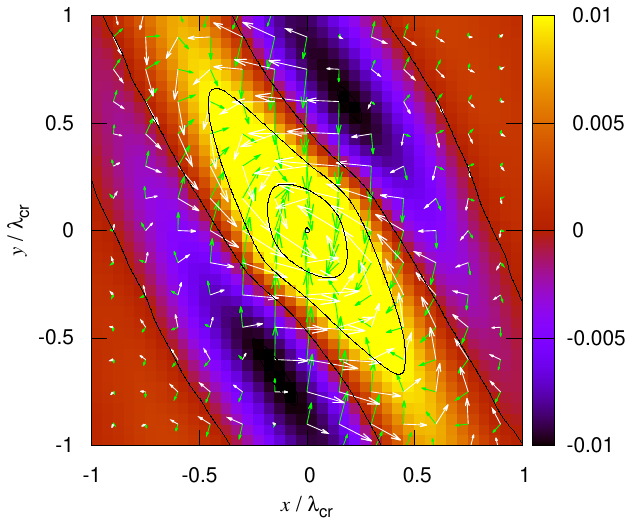} 
\end{minipage}
\begin{minipage}{0.49\hsize}
   \vspace{13pt}  
  \begin{center} $\Omega s/2\pi = 0.00$ \end{center}
  \vspace{-14pt}
  \plotone{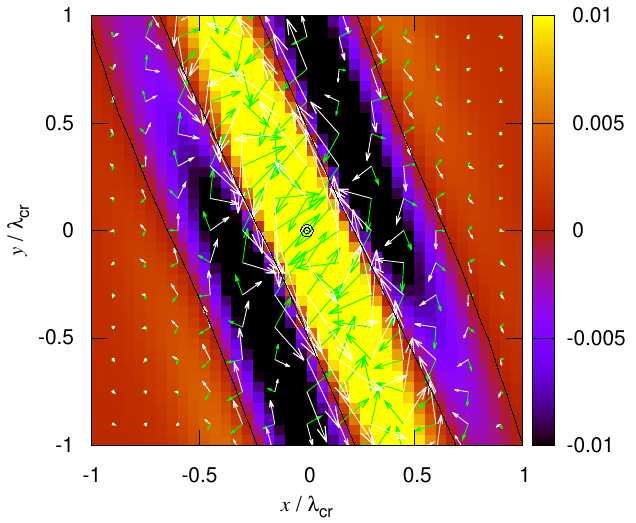} 
\end{minipage}
\begin{minipage}{0.49\hsize}
   \vspace{13pt}  
  \begin{center}  $\Omega s/2\pi = 0.40$ \end{center}
   \vspace{-14pt}
 \plotone{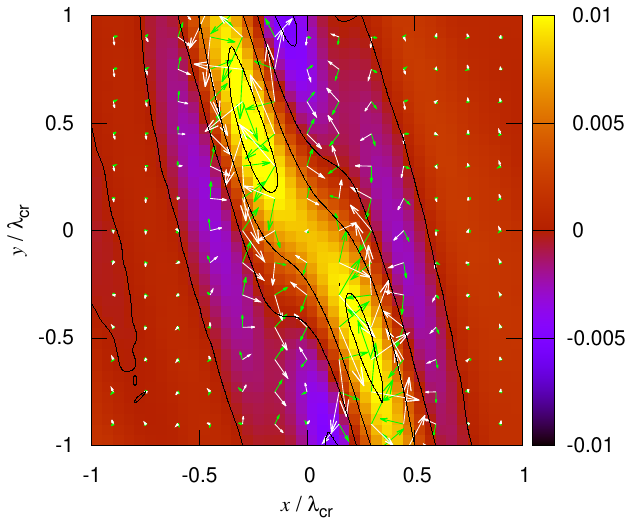} 
\end{minipage}
 \caption{
Cross-correlation function of the surface density
and the displacement from the guiding center (green arrows) and the relative velocity to the guiding center (white arrows) 
 at $\Omega s/2\pi = -0.40, -0.20, 0.00,$ and
 $0.40$ for model k4 (from top left to bottom right).
The color map denotes the space-time autocorrelation function of the surface
 density as in Figure \ref{fig:delyed_evolution}, and the solid curves denote its contours.
}
\label{fig:cross_cor}
\end{figure}

In order to elucidate the physical process in more detail, 
 we investigate the evolution of the displacement from the guiding
 center and the relative velocity to the guiding center using Equation (\ref{eq:crosscor}). 
We calculate the displacement vector from the guiding center by
 $\langle x-x_\mathrm{g} \rangle (x,y,s)$ and
 $\langle y-y_\mathrm{g} \rangle (x,y,s)$ and
 the relative velocity to the guiding center by
 $\langle v_x \rangle (x,y,s)$ and
 $\langle v_y+2Ax_\mathrm{g} \rangle (x,y,s)$, and show them in
Figure \ref{fig:cross_cor}. 
At $\Omega s/2\pi= -0.4$, we can see the leading structure. 
In this case, the rotation of the pattern cancels out that of the
 coordinate system. 
In the comoving frame of the leading pattern, the stabilizing effect due
 to rotation weakens. 
Thus, the stars are pulled towards the center of the density
 pattern, and the relative velocity is almost parallel to the leading 
 pattern. 
The stars move along the density pattern and the two density peaks come close to each other. 
Note that, this behavior is unpredictable by the analytical theory of the swing amplification. 
Since, in the analytical theory, we postulate the infinite plane wave that has no structure in the direction parallel to the wave,  the resulting perturbed flow towards the density peak is perpendicular to the wave. 
At $\Omega s/2\pi= -0.2$, the significant density enhancement takes place due
 to the flow towards the density peak.
Due to the Coriolis force, the velocity field rotates in the
 anticlockwise direction around the density peak. 
At the same time, the long trailing tail from the density peak appears.
At $\Omega s/2\pi= 0.0$, the clear trailing pattern appears.
The displacement from the guiding center has a convergent field towards
 the density peak. 
This indicates that the phase synchronization causes the density
 enhancement. 
On the other hand, the corresponding velocity field rotates in the
 anticlockwise direction around the density peak. 
Hence, around the density peak the region where $y > 0$ moves to the left and the region where $y < 0$ moves to the right.
At $\Omega s/2\pi= 0.4$, this anti-parallel motion splits the trailing
 pattern into two halves.

\section{Summary}

We have performed local $N$-body simulations of galactic spiral arms and investigated their amplification process in detail.
Using the time autocorrelation function, we estimated the typical
 lifetime of spiral arms.  
The dependence of the damping time of spiral arms on the epicycle frequency $\kappa$
 is consistent with the epicycle period. 
This indicates that the generation and destruction of spiral arms is
 ascribable to the epicycle motion.

In \cite{Michikoshi2016a} (Paper III), from the theoretical perspective, we pointed out that the phase synchronization of the epicycle motion would play an important role in the density amplification.
We investigated the spatial distribution of orbital elements of stars and found that the epicycle phase is synchronized in spiral arms while 
 the guiding center distribution is uniform (Figure~\ref{fig:snapshot_evo}). 

In order to understand the amplification mechanism in detail, 
we performed the delayed spatial autocorrelation analyses.
This shows the typical evolution of the surface density around the overdense regions.
The leading pattern appears before the density is amplified, which is consistent with the analytical theory.
In the leading pattern, stars move to the center because particles are pulled towards the center of the pattern.
Such a behaviour is not assumed in the analytical works based on the linear analysis because their analyses postulate a infinite plane wave.
The Coriolis force changes this convergent flow into the anticlockwise rotational flow. 
When the pattern is most amplified, the clear anticlockwise flow occurs.
Thus, in the pattern, anti-parallel flow arises, which splits the pattern into two halves. 
The results of $N$-body simulations indicate the importance of the finite length of spiral arms. 

As shown in Figures \ref{fig:delyed_evolution} and \ref{fig:cross_cor},
 two swarms of stars appear before and after the amplification. 
Thus, the basic picture of the amplification can be interpreted as the two-body interaction of the swarms. 
We consider two swarms whose galactocentric distances are different.
Due to the shear, they come close to each other.
As the distance between them becomes small, the self-gravity between
 them becomes strong. 
Then the epicycle motion is excited to approach to each other.
The swarms collide with each other and one large swarm forms.
The large swarm deforms to a trailing pattern with increasing the density. 
Because the system is collisionless, each swarm continues the epicycle 
 motion. 
After half an epicycle period, the two swarms separate away.
Then the spiral arms are destroyed.
The remaining swarms interact with another swarm and continue the spiral activity.  
This may be an elementary process for the formation of recurrent and
 transient spiral arms.  

\section*{Acknowledgement}

Numerical computations were carried out on GRAPE system at Center for Computational Astrophysics, National Astronomical Observatory of Japan.
S. Michikoshi is supported by JSPS KAKENHI Grant Number 17K14378.  E. K. is supported by JSPS KAKENHI Grant Number 18H05438.

\end{document}